\documentclass[letterpaper,twocolumn,10pt]{article}
\usepackage{usenix2019}

\usepackage[pdftex]{graphicx}
\graphicspath{{./Figures/}{./Figures/plots/}}

\usepackage{hyperref}
\usepackage{url}

\usepackage{amsmath}
\usepackage[utf8]{inputenc}
\usepackage{xspace}
\usepackage{hyphenat} 
\usepackage{comment}
\usepackage{booktabs}
\usepackage{subcaption}
\usepackage{multirow}
\usepackage{color, colortbl}
\usepackage{wasysym}
\usepackage{mdwlist}
\usepackage{cite}

\usepackage{balance}
\usepackage{soul}
\usepackage[anythingbreaks]{breakurl}
\usepackage[normalem]{ulem}
\usepackage[table,xcdraw]{xcolor}
\usepackage{rotating}
\usepackage{pgf}
\usepackage{afterpage}
\usepackage{pifont}

\definecolor{Gray}{gray}{0.9}

\usepackage{tikz}

\begin{document}
\title{Towards an Insightful Computer Security Seminar}

\author{
{\rm Kashyap Thimmaraju$^{1}$ \quad Julian Fietkau$^1$ \quad Fatemeh
Ganji$^{2}$}\\
\normalsize{{$^1$ Security in Telecommunications, Technische Universit{\"a}t Berlin \quad
 \quad $^2$ University of Florida
}}}

%
%
%
%

\maketitle

\subsection*{Abstract}
In this paper we describe our experience in designing and evaluating our
graduate level computer security seminar course. In particular, our seminar is
designed with two goals in mind. First, to instill critical thinking by teaching
graduate students how to read, review and present scientific literature. Second,
to learn about the state-of-the-art in computer security and privacy research by
reviewing proceedings from one of the top four security and privacy conferences including IEEE Symposium on
Security and Privacy (Oakland S\&P), USENIX Security,  Network and Distributed System Security Symposium (NDSS) and ACM Conference on Computer and Communications Security (CCS). The course entails each student to i) choose a
specific technical session from the most recent conference, ii) review and
present three papers from the chosen session and iii) analyze the relationship
between the chosen papers from the session. To evaluate the course, we designed
a set of questions to understand the motivation and decisions behind the
students’ choices as well as to evaluate and improve the quality of the course.
Our key insights from the evaluation are the following: The three most popular
topics of interest were Privacy, Web Security and Authentication, ii) ~33\% of
the students chose the sessions based on the title of papers and iii) when providing an encouraging environment, students
enjoy and engage in discussions. 

\sloppy

\section{Introduction}
As we undergo the digital transformation~\cite{digital-transformation}, education in computer security is
increasingly pertinent and challenging~\cite{bishop1993teaching,nsf-security,sherman2017insure}.
Graduate students need to be prepared for
their next step, be it academia or the industry.  Not only do they require a solid
foundation but they also need to be aware of the state-of-the-art. Oftentimes,
the later is not included in the curriculum for various reasons: It takes time
for textbooks to incorporate research from the previous months; research can be
eclectic and hence may not have a text-book; or educators find it overwhelming
to keep up with the vast amounts of new research and then include a subset in
their teaching material. Hence, seminars (or colloquiums) are an ideal setting
for students and researchers (graduate students, post-docs and professors) to
actively engage with the latest published research through presentations,
reading groups and discussions.

Widely adopted seminar syllabi are typically pedagogical in that, there are a
static set of seminal papers in computer security and privacy that students are
expected to study and review~\cite{dan,berk-security}. The class meets on a weekly
or biweekly basis and discusses the assigned paper. At the end of the semester
students might work on a project~\cite{stanford,paxson} and then submit a
technical report that summarizes the seminar topics~\cite{bonn} or the project.
Such a format exposes students to a solid overview of some of the first and
fundamental computer security contributions. 

However, there are a few disadvantages.  First, students do not get exposed to
the latest research contributions and developments, which has been growing at a
staggering rate: the top 4 security and privacy conference receive close to 3000
submissions a year (in total); and on average accept only around 16\% of them. Second, we
believe the seminar should be a setting where students can explore areas of
their interest given the broad spectrum of security research and technical
sessions in a conference.  
Third, compared to giving a presentation, handing in a write-up can be less preferred as the students often do not receive sufficient training in academic writing. Moreover, the short timeline of a colloquium does not allow the instructors to help the student with their writing skills. On the contrary, according to our observation, students can improve their presentation skills, when giving them feedback and comments in the class. They can further improve their communication skill as making constructive criticism and suggestions constitutes an important input to the evaluation. 


At our university, we have spent the last 4 (academic) years conducting the
computer security seminar with students pursuing a masters degree in Computer
Science. In this paper, we share our \emph{experience and feedback from
designing and evaluating our computer security seminar course}. In particular, we
believe our syllabus is novel and worth consideration by other educators around
the world for the following reasons.

First and foremost, students not only review papers from the latest conference proceedings
but also analyze the relationships between the papers in a technical
session from the conference. Usually, attendees and the wider audience of a
conference are not aware of the decisions behind the papers that make up a
session at a conference. In our seminar, each student attempts to understand not
only from a high level the connections between the papers in a seminar, but also
what makes a good session, or what was the key message of the session. For
example, one student reviewed 3 papers from the ``Deep Learning and Adversarial
ML'' session at NDSS 2018 and had the following takeaways (without any modification):``Neural Networks (NNs) perform well in many tasks, e.g., static code analysis, in particular for
finding vulnerabilities based on information flows analysis. However, NNs have
vulnerabilities, due to model differentiation even though some protection
mechanisms exist. Unfortunately, Models can't be verified (yet). And, don't use
untrusted models for critical tasks.''

Drawing such a conclusion is really helpful in quickly grasping what has been
published, but also in understanding what the bigger picture of the field is.
This can be valuable for researchers at different levels to find inspiration or
even build on existing work.
Moreover, although the conference proceedings are chosen by the instructors, the
students are free to choose the session, this makes the course more meaningful
to the students and instructors as everybody ``learns what they are
interested in''. We found that half of all the students who participated chose
their session because they were eager to learn more about the specific topic.
In addition to this, rather than have students write technical reports, instilling a sense of critical
thinking, scientific skepticism and presentation is actually appreciated by students and being valuable. We find this to be particularly useful in computer
security research with respect to threat models and evaluations.

\noindent\textbf{Organization.} %
The remainder of this paper is organized as follows. In
Section~\ref{sec:seminar}, we elaborate on the design goals and syllabus of our
seminar. Next, in Section~\ref{sec:outcomes} we describe the key outcomes and lessons
learned from several iterations of the seminar followed by a discussion of
related work in Section~\ref{sec:relatedwork}. Finally, in
Section~\ref{sec:conclusion} we conclude this paper.

\section{The Seminar}
\label{sec:seminar}
We first highlight what our goals were when we designed the seminar, and then
elaborate on the details of the course syllabus, grading and schedule.
\vspace{-1.0em}

\subsection{Design Goals}
\label{sec:design-goals}
We designed the seminar with the following goals in mind.
\begin{enumerate}
\item Expose students to the state-of-the-art in computer security and privacy, 
\vspace{-0.7em}
\item Expose students to a conference-like setting, 
\vspace{-0.7em}
\item Promoting critical thinking and scientific skepticism,
\vspace{-0.7em}
\item Provide an encouraging environment for students and instructors to have meaningful and
intellectual discussions related to security and privacy, 
\vspace{-0.7em}
\item Improve students' skills in reading, writing and criticizing scientific work as well as presenting scientific papers in a limited time, and
\vspace{-0.7em}
\item Improve students' communication skills by valuing constructive criticism and encouraging them to give feedback to a presenter. 
\end{enumerate}
The first goal is specific to computer security and privacy community whereas the others are expected to empower students to become a part of a larger and
more general research community.

\subsection{Syllabus}
\label{sec:syllabus}

Having outlined our design goals, we structured the seminar as follows. To
ensure high-quality discussions and interactions in an inspiring environment, we limit the number of
students to approximately 15.  Next, we (instructors) choose one or two of the most
recent conferences proceedings that are available online and ask the students to
choose a conference session, which typically consists of 3 papers. For example,
``Adversarial Machine Learning'' from the proceedings of CCS 2017~\cite{ccs2017-aml}.

After the students have chosen their conference sessions, we organize them into
groups of 3 and assign each group to present one of the following papers:
How to read a paper~\cite{keshav2007read}, Introduction~\cite{annesley2010cold},
Methods~\cite{annesley2010and}, Results~\cite{annesley2010show} and
Discussion~\cite{annesley2010discussion}.

We have selected these articles to help the students understand the general structure
of scientific papers and how to read a paper. By having them present
these preparatory papers, it allows us to give the students feedback on their
presentation skills, e.g., their ability to effectively communicate the key
ideas and concepts, delivering high quality presentations (slides and speech),
etc. We do not grade the students on these presentations as our intention in
this phase is to provide a learning environment for the students to
not only understand the challenges of presenting a paper, but also improve their skills for the final
presentation.

The students then have the rest of the academic semester to i) review the 3 papers from the session they have already chosen, and ii) prepare a 30-45
minute presentation on them. The presentation guidelines and review form are available
online~\cite{intro-slides, review}.  In this phase of the seminar, the students work on their tasks; however, they can contact the instructors if they face a challenge. The student are expected to read and comprehend new concepts as well as familiarize themselves with the related work to obtain a better
understanding of the area. 
Moreover, to excel in presenting a scientific topic, they are encouraged to watch publicly available presentations, etc.

One month before the final presentations, we offer the students a discussion
session, where they can ask us questions related to the presentation, the papers
they have chosen, so on and so forth. This typically helps the class to understand the
expectations. To help the students, we also share a few presentations we have deemed ``good''
and what is considered ``not good''.

For the final technical presentation scheduled similar to a one-week course, a week towards the end of the
semester, which is typically one or two weeks before the exams, is considered. According to our experience, the motivation level of the students is high during this time period, and therefore, we have chosen that. Depending on the teaching resources made available (e.g., room availability), we schedule
two to five talks per day. Indeed, this is at a smaller scale compared to a
conference style; however, this enables us to invest sufficient time and energy to have
constructive discussions and share feedback with the class and presenters. 
Based on the sessions selected by the students, we schedule the talks in the most
meaningful way. For example, in the summer semester of 2019, students, who chose papers from the sessions on Web
Security, Web Applications, Software Security presented on the
same day.

After the presentation, the students are given more time (usually, a week or two) to submit their
reviews of the papers via an online form. The form can be viewed online~\cite{review}. 
This is, of course, further helpful for students, who would serve on the program committee of conferences or as reviewers of journals. 
In addition, we also request the students to complete an optional feedback form
(also available online~\cite{feedback}), which can not only help to
improve the seminar repeatedly, but also serve as useful data to be shared with our community, as the goal of this paper is.

Upon receiving all the reviews, each instructor evaluates each review independently. Once this is completed, we discuss any
discrepancies in the grading of each student. In our experience, this can help
us avoid introducing bias and expedite the grading process.

\section{Outcomes and Lessons Learned}
\label{sec:outcomes}
When we held this seminar for the first time, only 5 students completed the seminar.  Due to the low turnout, the feedback we received was primarily oral.
In the semesters that followed, we ensured we recorded the feedback from the
students either in Likert scale or qualitative form.  Roughly speaking, 50\% of the
students have given us their feedback (the form can viewed online~\cite{feedback}).
Below, we describe the key takeaways and lessons learned from the past
four semesters of the seminar, in which in total, 57 students have
reviewed 171 papers.

\begin{figure}[t!]
    \centering
    \includegraphics[trim=0.0cm 0.0cm 0.0cm
0.0cm,clip=true,width=.49\textwidth]{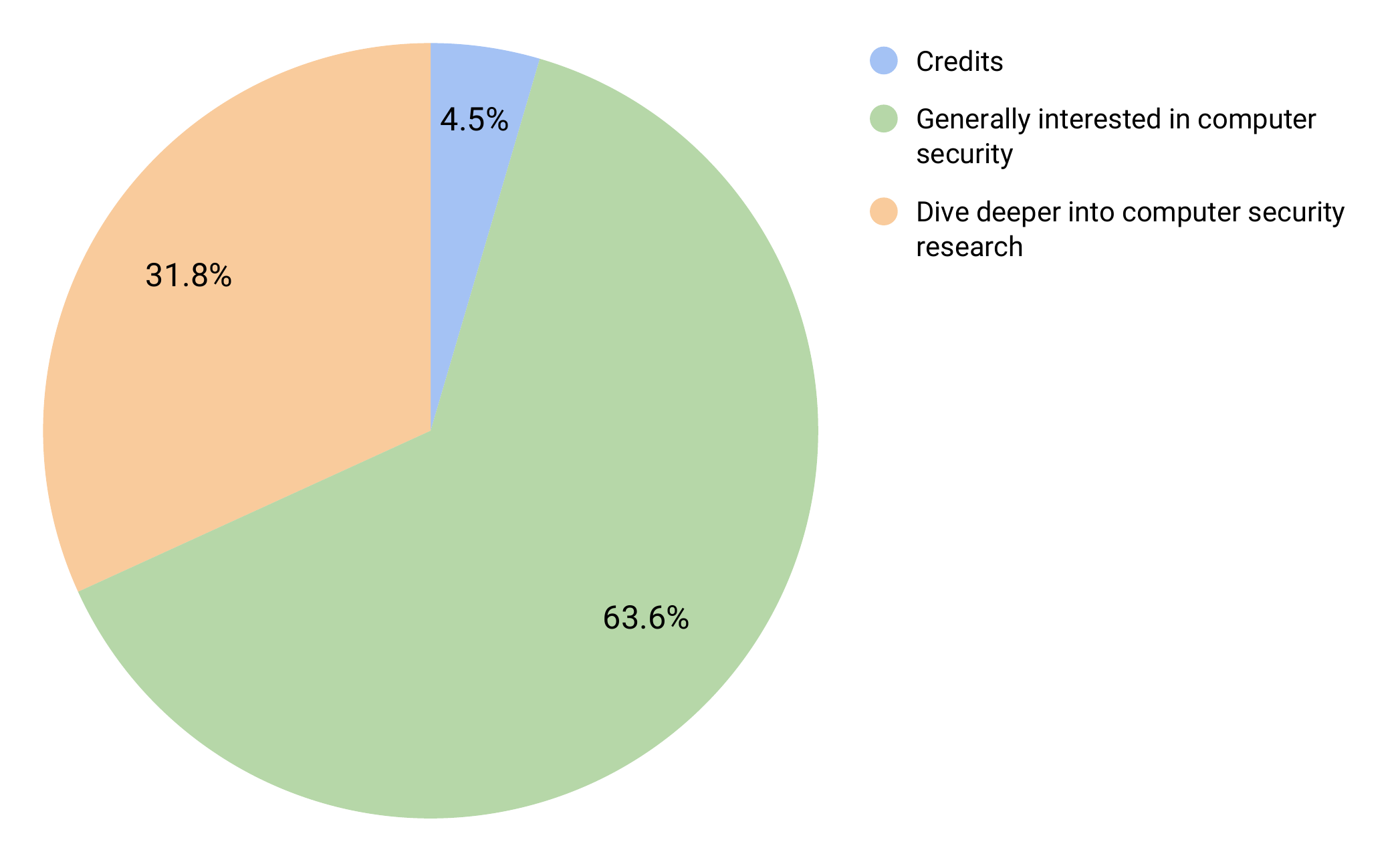}
	\caption{Pie chart of why students took the seminar.}
    \label{fig:whyseminar}
    \vspace{0.0em}
\end{figure}

\noindent\textbf{Choosing the seminar.} %
We observed that nearly all the students, who completed the seminar either were generally interested in computer security or planned to pursue computer
security-related research. Only one student took the course to earn the credits (see Figure~\ref{fig:whyseminar}). Furthermore,
nearly 35\% of the students were at an early stage in their master studies.  This
emphasizes the need to provide students (early on) with an environment, where
they can explore computer security-related topics and research based on their interests
rather than a set of prescribed topics. 
We observed that less than 10\% of the students planned to pursue research on security and 5\% took
the course as they would apply for a job in computer security industries.

\begin{figure}[t!]
    \centering
    \includegraphics[trim=0.0cm 0.0cm 0.0cm
0.0cm,clip=true,width=.49\textwidth]{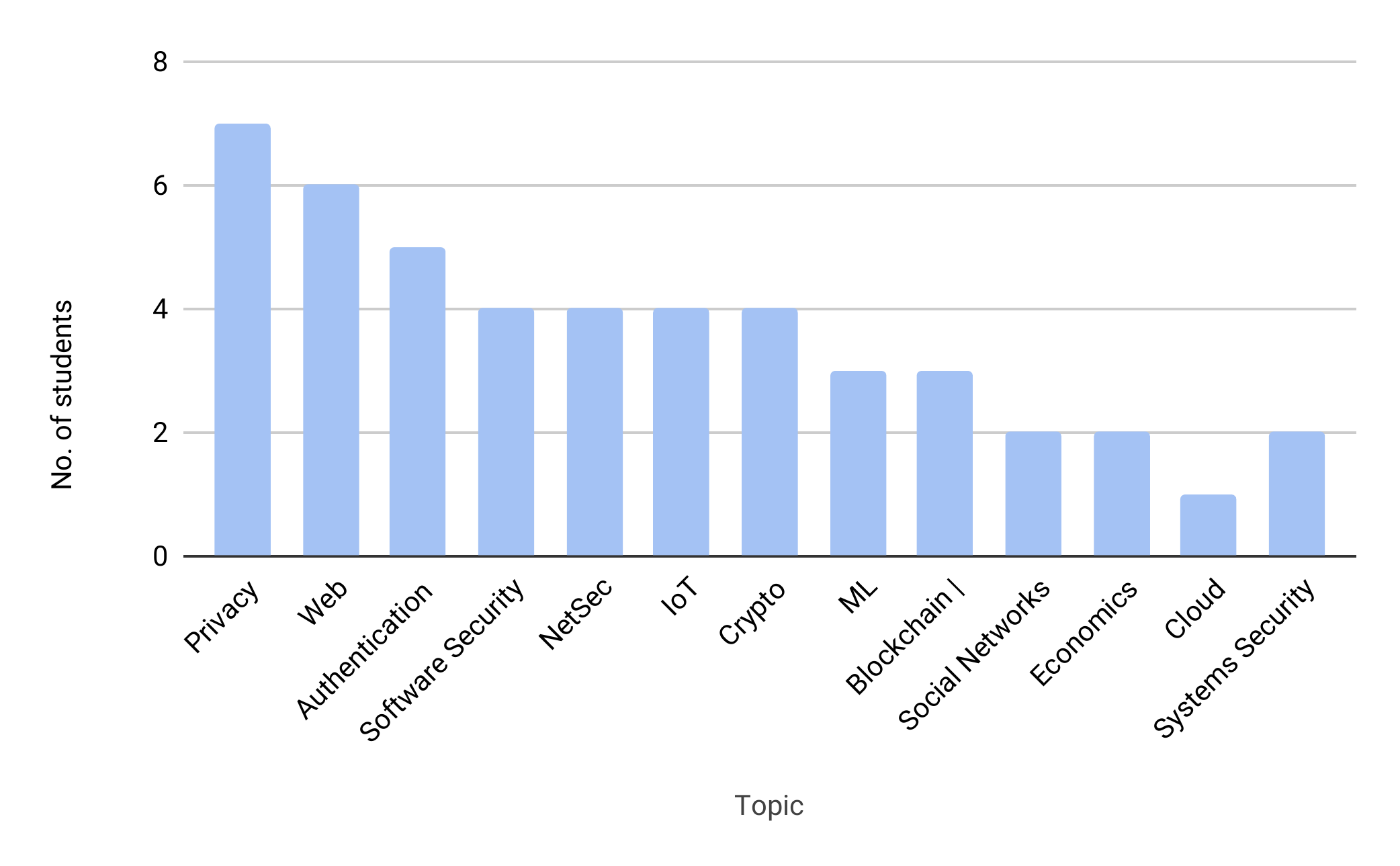}
	\caption{The distribution of topics chosen by the students from the
conferences.}
    \label{fig:topics}
    \vspace{0.0em}
\end{figure}

\noindent\textbf{Topics of interest.} %
The three most sought-after topics from the conference sessions were Privacy,
Web Security and Authentication. Software Security, Network Security, IoT and
Crypto followed in equal popularity as shown in Figure~\ref{fig:topics}. We observed that the topic ``Machine
Learning (ML) Security,'' which has been trendy in those four semester (Fall 2018-Spring 2019), was selected a few times by students who had sufficient
familiarity with ML. However, three students out of four who chose the Internet
of Things (IoT) sessions (from NDSS 2018 and CCS 2018), were not equipped with
the technical knowledge that the papers demanded, e.g., deep understanding of
software and systems security. For the most popular topics, we observed that
most students were able to grasp the technical concepts, however, only a handful
of students were capable of assimilating the deep technical details from the
papers.

\begin{figure}[t!]
    \centering
    \includegraphics[trim=0.0cm 0.0cm 0.0cm
0.0cm,clip=true,width=.49\textwidth]{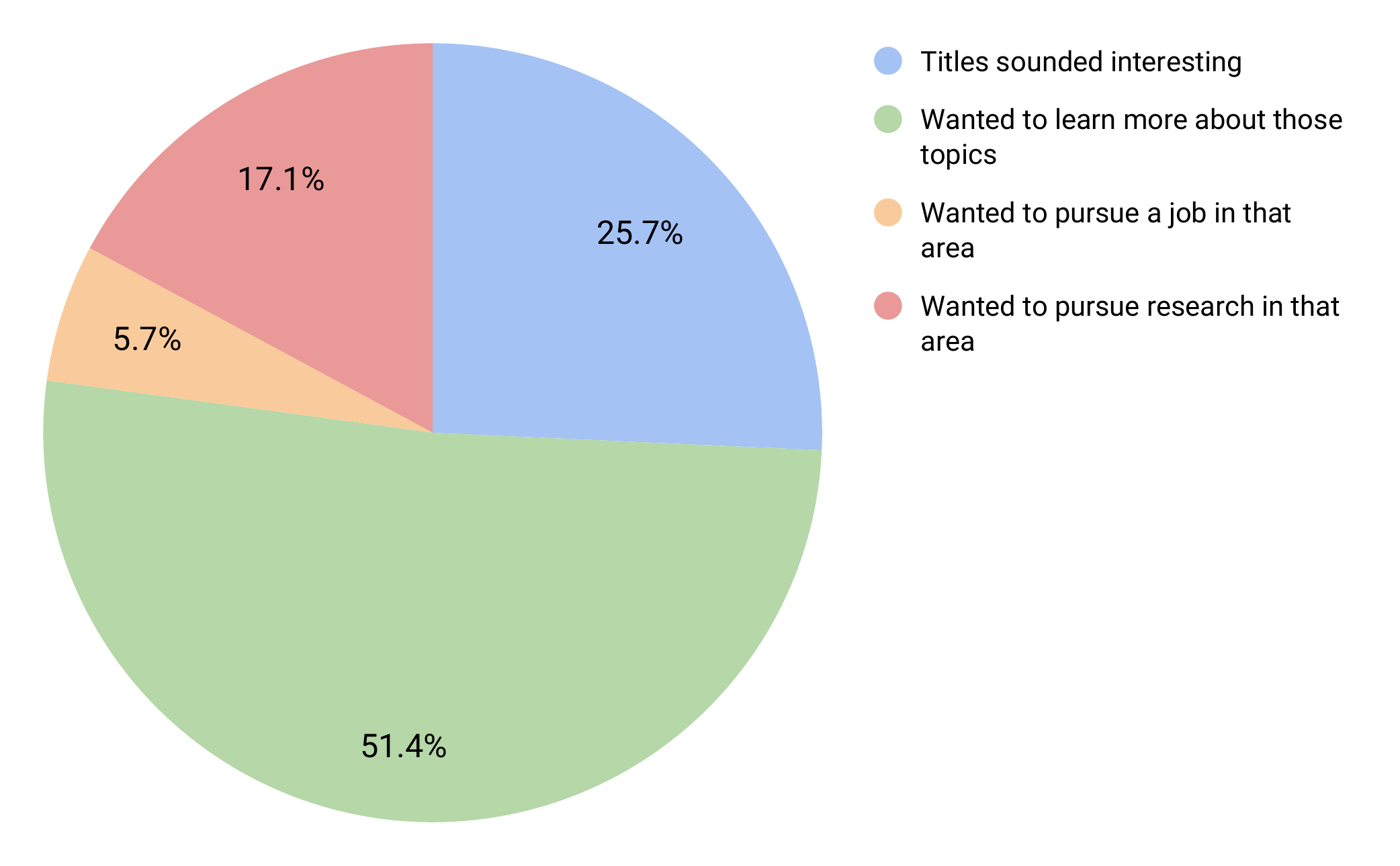}
	\caption{Pie chart of how students chose the conference session.}
    \label{fig:confsession}
    \vspace{-0.0em}
\end{figure}

\noindent\textbf{Choosing a session.} %
Depicted in Figure~\ref{fig:confsession}, approximately one third of the students choose their session based on the title of papers, which highlights the importance of a title to attract readers.  Nearly half of the
students aimed to learn more about specific topics, hence, chose sessions
accordingly. Very few students associated the session with their future, i.e., a
job in the industry or pursuing research. This points out that students attend the
seminar to explore the state-of-the-art in topics that pique their
interest. 
Note that in the conferences considered in our seminar, more than one session have been devoted to some topics. 
We also note that the students could choose more than one option here, therefore, the number of interested students is more than the total number of
participants.

\noindent\textbf{Preparatory papers.} %
Students found the set of articles, given to them as introductory reading material, helpful. In fact, students, who were
initially skeptical of the usefulness of those articles, found them helpful in reviewing the
papers chosen by them. 

\noindent\textbf{Student takeaways.} %
From the feedback we collected, there were three main benefits for the students.
The first was being positively surprised by the diversity, complexity and difficulty of
computer security. Second, they learned how to critically read, review and
present scientific literature. Third, it helped improve their presentation and
critical thinking skills in English (as many students are not native English
speakers). 

\noindent\textbf{Paper summaries.} %
The final review paper handed in by students include a summary of each paper in the session chosen by them. 
This paper has offered us (instructors) an overview of the latest methods, systems and problems
addressed. Furthermore, we believe it enables us to analyze trends across
conferences within an academic year as well as across years. We are currently
working on a collation of the summaries from the papers reviewed in our seminar, 
which covers sessions from conferences in 2018 and 2019.

\noindent\textbf{Reviewing a paper.}
In the first semester, when we held this seminar, the final review paper \emph{summarized} three papers in the session chosen by each student. We then realized that this could not align well with our design goals.
Hence, we switched to a format where the students (critically) review
the papers in their session. This was well received not only by the
students, but also by us for various reasons. First, it promotes critical thinking about scientific papers. 
Besides, for the student, reviewing a paper can further add a sense of engagement in the community. 
Second, the review papers could become concise and to the point.  Third, it gives the
students a sense of the real world paper submission experience. Finally, by
asking the students to describe the relationship between the papers in the
session, they can better grasp the notion of interdisciplinary research studies coming together under the umbrella of ``a conference session.'' 
To aid the students in these processes, we shared several
resources with them, e.g., reviews we received from conference submissions, how
to review a paper~\cite{keshav2007read}, Benchmarking crimes~\cite{van2018benchmarking}, Security
Experiments~\cite{peisert2007design}, etc.

\noindent\textbf{Benefits to instructors.} %
After instructing this seminar for four semester, we have obtained several benefits. First, the presentations and paper summaries have given us a holistic view of the state-of-the-art from the top security conferences.
Second, we have been able to find excellent students to assist us in our
research that has resulted in theses and publications. Third, the discussions
and talks have led to new ideas through the critical analysis of papers
that appear in the proceeding of top-tier conferences. Finally, we have also improved our own teaching style. 




\noindent\textbf{Students' feedback.}
Overall, the majority of the feedback from our students was positive. 
One of their comments on the seminar schedule was to hold weekly meetings instead of the block style. 
Regarding improvement, we were requested to give student more time to present details provided in the paper. 
One the one hand, it shows how interested they are in the topics offered in the seminars. 
Whereas, one the other hand, it reflects the fact that other lectures should be developed to cover the details of the new approaches published recently. 
This becomes more evident as covering such details does not fall within the scope of our seminar, as summarized under our design goals, see~\ref{sec:design-goals}. 
Here we quote some of the positive and negative feedback we received from the students (without any modification). 

``Definitely keep the format of reading 3 (or at least multiple) papers, since
this gives us some room to discuss the topics in a broader context than simply
presenting the contents of a single paper. I also really enjoyed that we had the
possibility and were encouraged to give our own opinions and criticism, during
the presentations as well as afterwards in the discussion. I think I learned a
lot more than in previous seminars where I only had to repeat the contents of a
paper without having to do a lot of thinking myself.''

``Good job! Especially the discussion and feedback after each presentation were a
good idea. I also like the format: Papers from RECENT TOP conferences (not some
unknown texts which have almost the same age as me) are presented and reviewed.
This is especially helpfull, for the move from master studies to doing research
as phd student.''

``Usually, I just read the useful parts of the paper to help me write the paper,
but this time, I read the whole articles and let me fully understand the way of
solving the problem. On the other hand, the topics are also interesting, I
have learned a lot of new ideas.''

``Maybe add a few notes on what exactly is expected in the review form. How much
text is required or how detailled the explanations are supposed to be. Maybe be
clear about the attendance up front.''

\section{Related Work}
\label{sec:relatedwork}
We have surveyed computer security seminars offered at other universities
and observe that various seminar instructors do indeed have students review papers from
conference
proceedings~\cite{dan,gottingen,formal,jaws,stanford,decentral,utah,dan}.
Furthermore, similar to our setting, some instructors even provide their students with helpful resources
and tips on how to read, review, write and present scientific
work~\cite{bonn,utah}. Oftentimes, the papers that students are suggested to
read are selected by the instructor/professor~\cite{decentral,jaws,wfu,stanford}, which mainly include seminal papers and project work~\cite{paxson}. In one
instance~\cite{orr}, we found the seminar involves students presenting chapters
from Ross Anderson's text book~\cite{anderson2008security}.  Our syllabus rather
focuses on various technical sessions from at least one or two recent conferences. 

\section{Conclusion}
\label{sec:conclusion}
In this paper, we described the design goals and structure of our computer security
seminar. We also discussed the key outcomes and lessons learned after instructing this seminar for 2 academic
years. In particular, we believe that our seminar syllabus offers students an encouraging environment to explore and learn the latest approaches and contributions in computer security and privacy community.  Furthermore, the
review process established by us, especially, the emphasis placed on the relationships between papers in a conference
session, instills a sense of confidence in students and promotes critical thinking.  In particular, constructive criticism is
necessary and valuable for students to improve their skills, essential for their future studies, research and jobs. 

\section{Acknowledgments}
We express our sincere thanks to Jean-Pierre Seifert who encouraged and
supported us in designing this seminar. We also thank our colleagues at
the Chair for Security in Telecommunications, Technische Universität
Berlin for their feedback and criticism when we designed and evaluated the seminar.
A special thanks to all the students who participated in the seminar
as well as for sharing their feedback with us.
{
{
\bibliographystyle{unsrt}
\small
\bibliography{cset2020}}
}


%
%



\end{document}